\documentclass[]{pasj01}
\draft

\Received{}
\Accepted{}
 
\begin{document} 

\title{ 
Rapid and efficient mass collection by a supersonic cloud-cloud collision as a major mechanism of high-mass star formation}

\author{Yasuo \textsc{Fukui}\altaffilmark{1}}%
\altaffiltext{1}{Department of Physics, Nagoya University, Furo-cho, Chikusa-ku Nagoya, 464-8602, Japan}
\email{fukui@a.phys.nagoya-u.ac.jp}

\author{Tsuyoshi \textsc{Inoue}\altaffilmark{1}}
\email{tsuyoshi.inoue@nagoya-u.jp}

\author{Takahiro \textsc{Hayakawa}\altaffilmark{1}}
\email{t.hayakawa@a.phys.nagoya-u.ac.jp}

\author{Kazufumi \textsc{Torii}\altaffilmark{2}}
\altaffiltext{2}{Nobeyama Radio Observatory, National Astronomical Observatory of Japan (NAOJ),
National Institutes of Natural Sciences (NINS), 462-2, Nobeyama, Minamimaki, Minamisaku,
Nagano 384-1305, Japan}
\email{kazufumi.torii@nao.ac.jp}


\maketitle

\begin{abstract}
A supersonic cloud-cloud collision produces a shock-compressed layer which leads to formation of high-mass stars via gravitational instability.
We carried out a detailed analysis of the layer by using the numerical simulations of magneto-hydrodynamics which deal with colliding molecular flows at a relative velocity of 20\,km\,s$^{-1}$ \citep{2013ApJ...774L..31I}.
Maximum density in the layer increases from 1000\,cm$^{-3}$ to more than $10^{5}$\,cm$^{-3}$ within 0.3\,Myrs by compression, and the turbulence and the magnetic field in the layer are amplified by a factor of $\sim 5$, increasing the mass accretion rate by two orders of magnitude to more than $10^{-4}$\,$M_{\solar}$\,yr$^{-1}$.
The layer becomes highly filamentary due to gas flows along the magnetic field lines, and dense cores are formed in the filaments.
The massive dense cores have size and mass of 0.03 -- 0.08\,pc and 8 -- 50\,$M_{\solar}$ and they are usually gravitationally unstable.
The mass function of the dense cores is significantly top-heavy as compared with the universal IMF, indicating that the cloud-cloud collision triggers preferentially the formation of O and early B stars.
We argue that the cloud-cloud collision is a versatile mechanism which creates a variety of stellar clusters from a single O star like RCW120 and M20 to tens of O stars of a super star cluster like RCW38 and a mini-starburst W43.
The core mass function predicted by the present model is consistent with the massive dense cores obtained by recent ALMA observations in RCW38 \citep{2019arXiv190707358T} and W43 \citep{2018NatAs...2..478M}.
Considering the increasing evidence for collision-triggered high-mass star formation, we argue that cloud-cloud collision is a major mechanism of high mass star formation.
\end{abstract}

\section{Introduction}\label{sec:introduction}
Formation of high-mass stars has been an issue of compelling interest in the last two decades because the influence of high-mass stars in galactic evolution is overwhelming and profound.
Considerable efforts have been paid toward understanding high mass star formation by a number of papers \citep{2014prpl.conf..149T,2015ASSL..412...43K}.

Three scenarios have been considered until now in the literature.
They are (1) stellar collisions in a rich stellar cluster \citep{1998MNRAS.298...93B}, (2) the competitive accretion in a massive self-gravitating system \citep{2001MNRAS.323..785B}, and (3) the monolithic collapse of a dense cloud \citep{2002Natur.416...59M,2009Sci...323..754K}.
If the stellar density is high enough for frequent collision between stars, collisional merging of low-mass stars may work as a possible mechanism.
Later, it was proved that the stellar density has to be uncomfortably large for such collisions to be efficient, and stellar collisions is no more considered as a viable mechanism of high mass star formation \citep{2011MNRAS.410.2799M}.
The other two mechanisms remain to be confronted by observations, and it is not yet settled how high-mass stars are forming (see for a review \cite{2014prpl.conf..149T}).
A possible argument unfavorable to the competitive accretion is an isolated O star such as the exciting star of M20 and RCW120, which does not belong to a massive cluster (\cite{2011ApJ...738...46T,2015ApJ...806....7T}; see also \cite{2018ASSL..424....1A}).
The numerical simulations of the monolithic collapse by \citet{2009Sci...323..754K} showed the formation of O stars from molecular gas of 100\,$M_{\solar}$ confined within 0.1\,pc.
This initial condition was adopted from observations of a protostar candidate IRAS 05358+3543 in the S233 -- 235 high-mass star forming region and other typical regions of high-mass star formation (\cite{2007A&A...466.1065B}; see also \cite{2002Natur.416...59M}).
A recent analysis of molecular gas by \citet{2020arXiv200110693K} showed that IRAS 05358+3543 was possibly formed by an external trigger of a cloud-cloud collision, an ad-hoc situation externally driven, raising caution on the initial condition adopted by \citet{2009Sci...323..754K}. 

High-mass star formation requires accumulating large cloud mass into a small volume.
Suppose that the initial condition is a massive low-density cloud and the final phase is a small dense cloud of the same mass which rapidly forms high-mass star(s).
This scenario requires an intermediate phase when cloud density is high enough to form low-mass stars unless the phase has a time scale much shorter than the free-fall time scale.
During the phase star formation has to be strongly suppressed.
Otherwise, the cloud loses mass by star formation before being collected into the small volume and high-mass star(s) cannot be formed.
Such a naive thought may be useful to shed light on the essence of the high-mass star formation as suggested in a review article by \citet{2007ARA&A..45..481Z},
\begin{quote}
``Rapid external shock compression (i.e., supersonic gas motions) generating high column densities in less than a local free-fall time rather than slow quasistatic build-up of massive cores may be the recipe to set up the initial conditions for local and global bursts of massive star formation''
 (in Section 9 of \cite{2007ARA&A..45..481Z}), 
\end{quote}
while no further pursuits on ``the rapid external compression'' are found in the literature. 

Cloud-cloud collision is another promising mechanism of the high-mass star formation as shown by recent theoretical and observational works, and has become a possible alternative to the previous scenarios of high-mass star formation above, (2) and (3).
\citet{1992PASJ...44..203H} showed that, in a supersonic collision between small and large molecular clouds, the small cloud creates a cavity in the large cloud, and the layer between the two clouds are strongly compressed to higher density, where a condition favorable for high-mass star formation is realized \citep{1992PASJ...44..203H,2010MNRAS.405.1431A,2014ApJ...792...63T,2018PASJ...70S..54S}.
After the onset of the collision, the two clouds show complementary distribution between the cavity and the small cloud, and the interface layer shows a broad bridge feature connecting the two clouds in velocity \citep{2015MNRAS.454.1634H}.
Recent observations show that in more than 50 regions of high-mass star formation these observational signatures of collision are identified and typical cloud properties are derived such as density and mass of the colliding clouds \citep{2019PASJ..tmp..127E}.
These objects includes HII regions and super star clusters in the Milky Way and the Local Group (e.g., M42/M43 \cite{2018ApJ...859..166F}; M17 \cite{2018PASJ...70S..42N}; NGC6334 and NGC6357 \cite{2018PASJ...70S..41F}; M20 \cite{2011ApJ...738...46T,2017ApJ...835..142T}; RCW120 \cite{2015ApJ...806....7T}; RCW38 \cite{2016ApJ...820...26F}; Westerlund2 \cite{2009ApJ...696L.115F,2010ApJ...709..975O}; and NGC3603 \cite{2014ApJ...780...36F}; R136 \cite{2017PASJ...69L...5F}).

It is important to understand the detailed properties of the forming stars under a trigger of a cloud-cloud collision, and theoretical understanding of the physical properties of the colliding clouds is crucial.
Magneto-hydrodynamic (MHD) numerical simulations were made for molecular gas flows colliding supersonically in a plane-parallel configuration by \citet{2013ApJ...774L..31I}, and showed that dense cores around 100\,$M_{\solar}$, precursors of high-mass stars, are formed.
These simulations showed that the layer is characterized by enhanced turbulence and magnetic field where the effective sound velocity is increased by several times depending on the collision velocity.
As a result, a mass accretion rate, which is proportional to the third power of the effective sound speed, becomes as high as $10^{-4}$ -- $10^{-3}$\,$M_{\solar}$\,yr$^{-1}$, which satisfies a value high enough to overcome the stellar radiation feedback \citep{1987ApJ...319..850W}.
Since the simulations include more details of the compressed layer such as the mass function of the dense cores, which were not presented by \citet{2013ApJ...774L..31I}, it is worthwhile to explore further the molecular gas properties in the simulations. Such a study will shed light on the time evolution in a cloud-cloud collision and help us to obtain an insight into the triggered high-mass star formation.

The aim of the present paper is to analyze the simulation data of \citet{2013ApJ...774L..31I} and to explore the details of the gas dynamics and the collision products.
In particular, we aim to derive a core mass function in a cloud-cloud collision and their detailed physical properties, which will allow us to make a comparison between the theory and the observations, and deepen our understanding of the role of cloud-cloud collision.
This paper is organized as follows.
Section \ref{sec:model} describes the model of \citet{2013ApJ...774L..31I} and Section \ref{sec:compressed_layer} the outcomes of the simulations.
Section \ref{sec:properties_cores} gives the dense core properties including the distribution of the cores and the core mass function.
Confrontations with the observations are given in Section \ref{sec:comparison_with_obs}.
Section \ref{sec:model_dependency} discuss the dependence on the model parameters, and Section \ref{sec:conclusions} gives conclusions.

\section{The model}\label{sec:model}
\begin{figure}
 \begin{center}
  \includegraphics{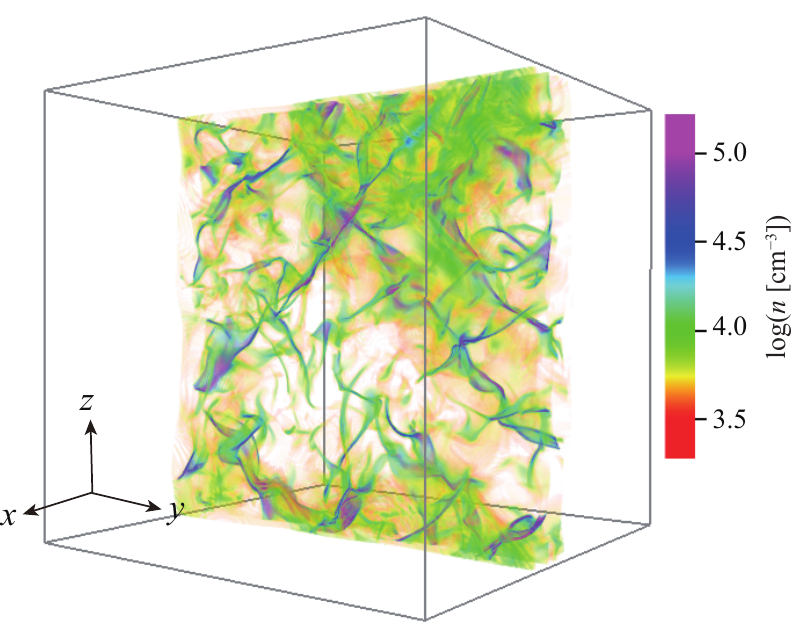} 
 \end{center}
\caption{Volume rendering map of density at $t=0.6$\,Myr (reproduced from \cite{2013ApJ...774L..31I}).}\label{fig:volume_render}
\end{figure}

Figure \ref{fig:volume_render} shows a three-dimensional view of the shock compressed layer in a volume rendering map at an epoch of 0.6\,Myrs after the onset of the collision, where the three axes $x$, $y$ and $z$ are shown in a simulation box, which has a size of 8\,pc in each axis and the pixel size is $8\,\mbox{pc}/512=0.015\,\mbox{pc}$ or 3000\,AU.
Two colliding molecular flows along the $x$ axis at 10\,km\,s$^{-1}$ in the box frame create the shock compressed layer in the central $y$-$z$ plane of the box.
The simulations were made in the ideal MHD with self-gravity.
The molecular cooling time scale is less than $10^{4}$\,yrs and the gas is approximated to be isothermal at 12\,K, corresponding to a sound speed of 0.2\,km\,s$^{-1}$.
The initial magnetic field is taken to be 20\,$\mu$G and the field direction is chosen to be perpendicular to the collision velocity.
The assumption does not substantially affect the final outcome because the colliding flows strongly compress the magnetic field lines in a direction perpendicular to the collision velocity.
The initial gas density is 300\,cm$^{-3}$ with a density fluctuation of $\sim30$\%.
The model setup is summarized in Table \ref{tab:modelparameters} and more details are given in \citet{2013ApJ...774L..31I}.
The calculations are made up to 0.7\,Myrs and the numerical results are recorded every 0.1\,Myrs.

\section{Structure of the shock compressed layer and the formation of the dense cores}\label{sec:compressed_layer}

\subsection{The shock compressed layer}
\begin{figure*}
 \begin{center}
  \includegraphics{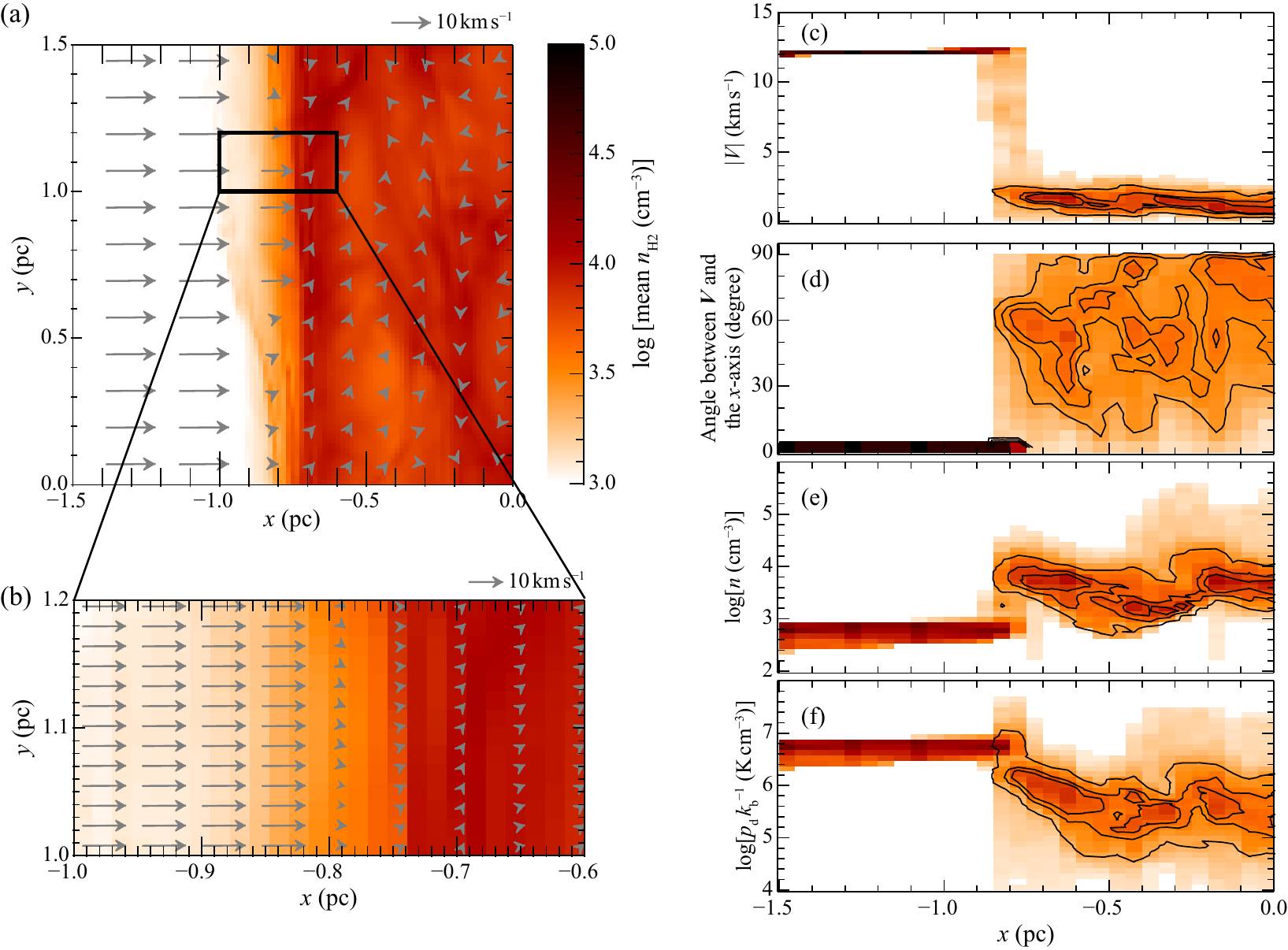} 
 \end{center}
\caption{(a) Projected velocity vectors (arrows) overlaid on distribution of mean density (image) in $|z|<0.08$\,pc at $t=0.7$\,Myr.
(b) Close up view of (a).
(c) Distribution of size of velocity vector ($|V|$), (d) angle between velocity vector and the $x$-axis, (e) density $n$ and (f) dynamic pressure for each pixel.
The contours in panels (c) -- (f) contain 30, 60, and 90\% of data points but excluding gas with initial condition ($|V|\gtsim 12$\,km\,s$^{-1}$ and angle $\sim 0\degree$).
}\label{fig:velocity}
\end{figure*}

The colliding gas has inhomogeneous density distribution.
The flows create the compressed layer between them, which becomes decelerated and denser.
Figure \ref{fig:velocity}(a) shows the velocity vectors in an $x$-$y$ plane on the negative-$x$ side, and the behavior is similar on the positive $x$ side.
Figure \ref{fig:velocity}(b) shows a close up in the velocity transition zone at an epoch of 0.7\,Myr.
The velocity does not change from the initial value at $x$ less than $-1.0$\,pc, and quickly becomes small and completely randomized in angle at $x=-1.0$\,pc -- $-0.6$\,pc.
This change is due to the shock dissipation and momentum exchange between the colliding flows.
Figures \ref{fig:velocity}(c) -- (f) show the averaged velocity, the angle between the velocity vector and the $x$ axis, the gas density and pressure respectively, as a function of $x$, and show that the velocity decreases to $\sim 2$\,km\,s$^{-1}$ from 12\,km\,s$^{-1}$ and is randomized in direction.
In Figures \ref{fig:velocity}(a) and 2(b) velocity is larger than the initial value 10\,km\,s$^{-1}$ because of the gravitational acceleration by the shock compressed layer.

\begin{figure}
 \begin{center}
  \includegraphics{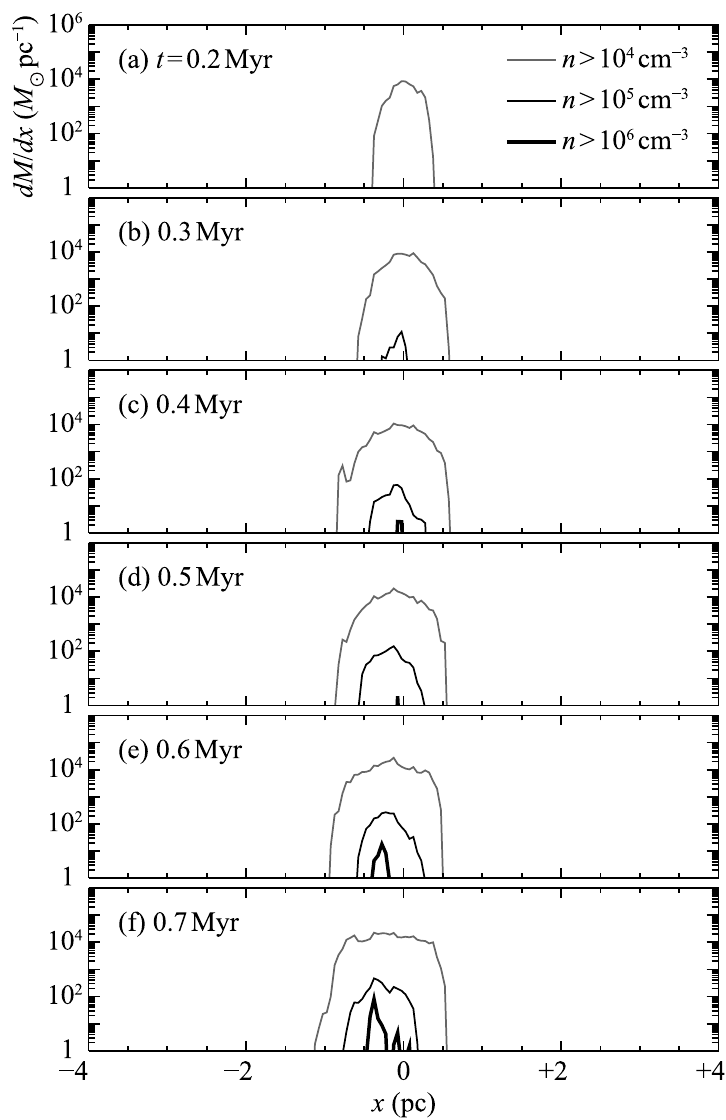} 
 \end{center}
\caption{
(a) Distribution of dense molecular gas along the $x$-axis at $t=0.2$\,Myr.
(b) -- (f) Same as (a) but for $t=0.3$--0.7\,Myr.
Gray, thin-black and thick-black lines in each panel show $n>10^{4}$\,cm$^{-3}$, $n>10^{5}$\,cm$^{-3}$ and $n>10^{6}$ cm$^{-3}$, respectively.
}\label{fig:density_ditribution}
\end{figure}

\begin{figure}
 \begin{center}
  \includegraphics{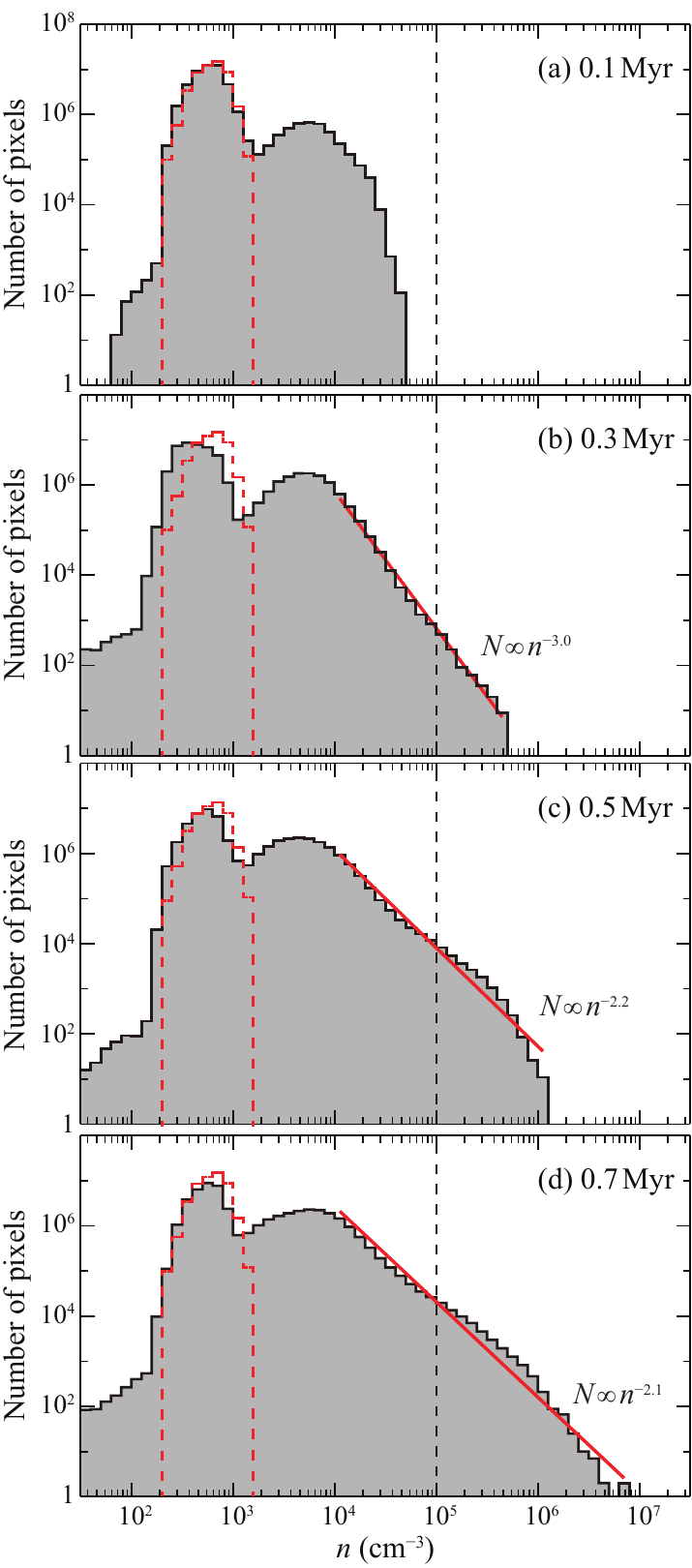} 
 \end{center}
\caption{(a) Histogram of H$_{2}$ density for each pixel in a range of $-1.5\,\mathrm{pc} <x<+1.5\,\mathrm{pc}$ at $t=0.1$\,Myr.
Histogram at $t=0$ is also shown by red dashed-line.
(b) -- (e) Same as (a) but for $t=0.1$, 0.3, 0.5 and 0.7\,Myr.
Vertical dashed line in each panel shows $n=10^{5}$\,cm$^{-3}$.}\label{fig:histo_density}
\end{figure}

The time evolution of the layer is shown as a density distribution along the $x$-axis at each epoch in Figure \ref{fig:density_ditribution}.
In 0.3\,Myr the maximum density becomes more than $10^{5}$\,cm$^{-3}$ and the fraction of the dense gas increases in time.
Table \ref{tab:mass} shows the gas mass in the compressed layer above $10^{4}$\,cm$^{-3}$.
The full thickness of the compressed dense layer above $10^{4}$\,cm$^{-3}$ is $\sim 1.5$\,pc in the $x$ direction.
After 0.4\,Myr we find sharp narrow spikes which indicate formation of filaments and dense cores in the layer as detailed later.
The fraction of the dense gas above $10^{5}$\,cm$^{-3}$ is 10\% -- 17\% of the total gas mass of density above $10^{4}$\,cm$^{-3}$.
Figure \ref{fig:histo_density} shows a probability distribution function of density, where $t=0.0$\,Myr shows the initial condition.
The collision creates the higher density tail above $10^{4}$\,cm$^{-3}$ as well as the low-density tail below the initial density distribution produced by turbulence.
The high-density tail characterizes the compression as expressed by a power law with an index of $-3.0$ -- $-2.0$, which becomes flatter in time.

\subsection{Dense cores in the filaments}
\begin{figure*}
 \begin{center}
  \includegraphics{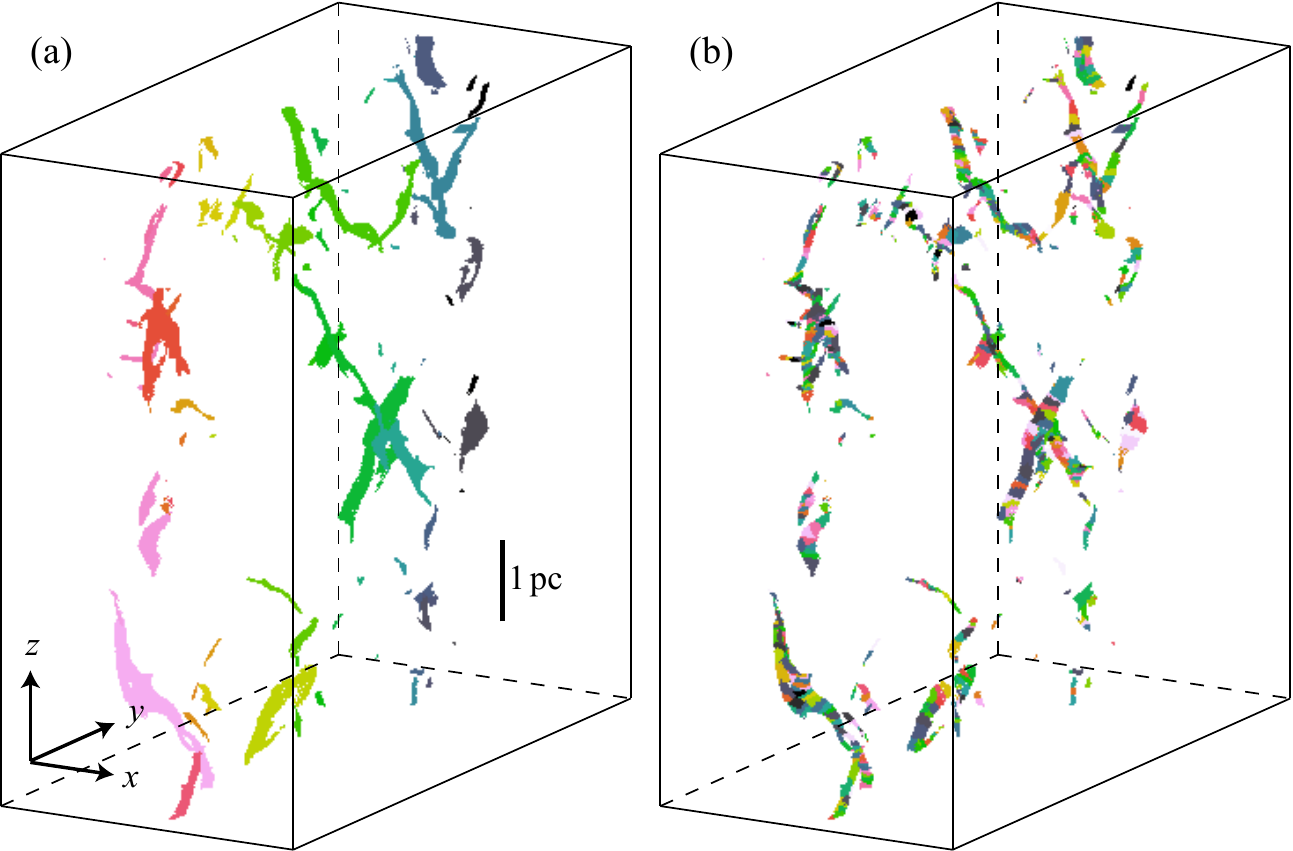} 
 \end{center}
\caption{
The distribution of (a) filamentary distribution, identified as connected region with $n>10^{5}$\,cm$^{-3}$ and (b) dense cores defined by using CLUMPFIND algorithm \citep{1994ApJ...428..693W}. 
Different filaments/cores in panels (a) and (b) are indicated by different arbitrary colors.
}\label{fig:filaments+cores}
\end{figure*}

The density distribution in the three-dimensional space is characterized by filaments which include dense cores.
Figure \ref{fig:filaments+cores}(a) shows the distribution of filamentary distribution identified as connected regions with $n>10^{5}$\,cm$^{-3}$, where different filaments are indicated by different arbitrary colors.
In each filament we define dense cores by using CLUMPFIND algorithm \citep{1994ApJ...428..693W}.
The algorithm is designed to be applied to spectral line datasets, $T(x, y, V)$ data cubes, but here we use it to find cores in $n(x, y, z)$ datasets.

\begin{figure*}
 \begin{center}
  \includegraphics{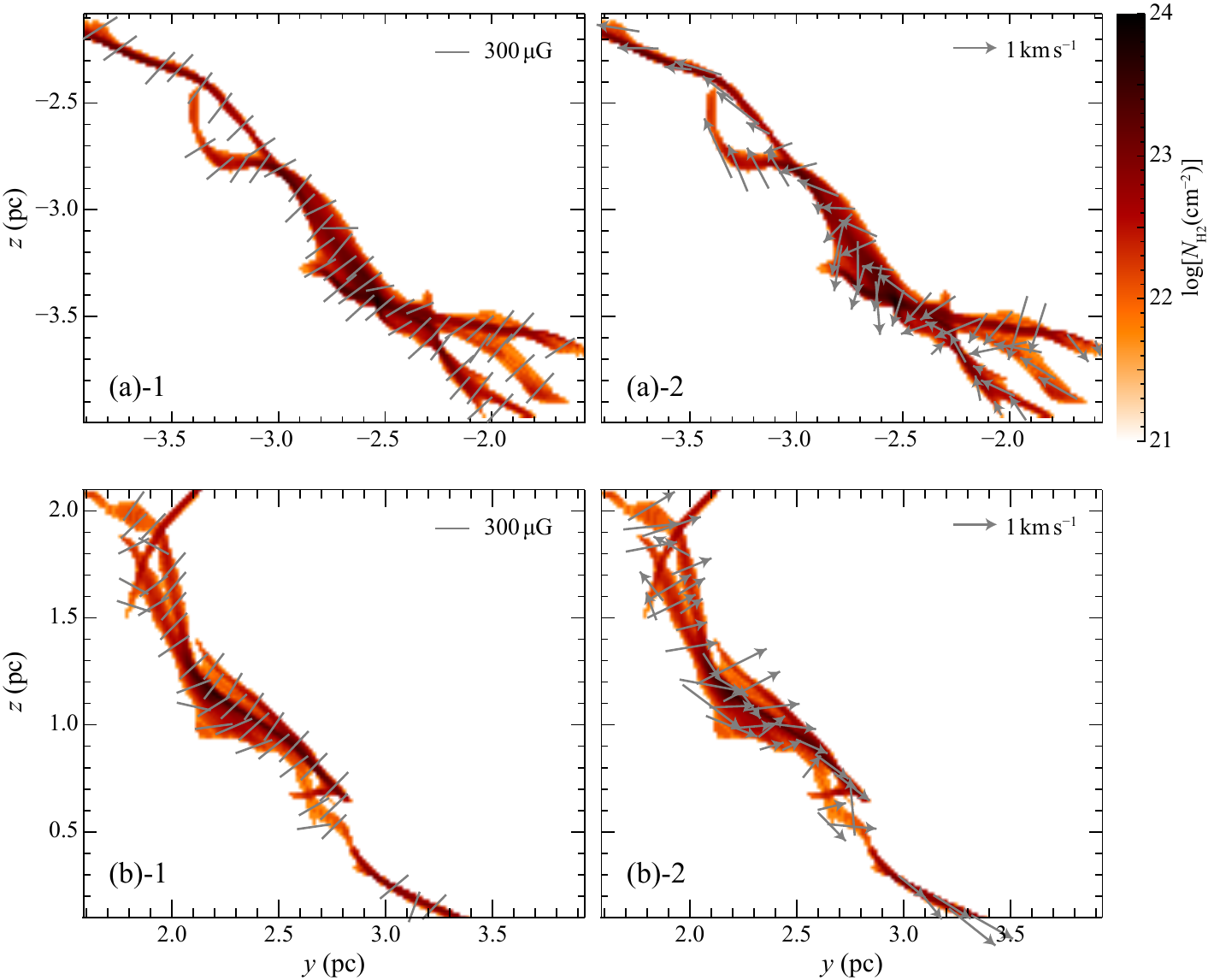} 
 \end{center}
\caption{
(a) Projected magnetic field vector ((a)-1) and velocity vector ((a)-2) shown in a typical region with filaments.
The magnetic field vector and velocity vector are given as $\mbox{\boldmath $B$}=\sum_\mathrm{filament}\left[\mbox{\boldmath $B$}(x)n(x)\Delta x\right]/\sum_\mathrm{filament}\left[n(x)\Delta x\right]$ and $\mbox{\boldmath $V$}=\sum_\mathrm{filament}\left[\mbox{\boldmath $V$}(x)n(x)\Delta x\right]/\sum_\mathrm{filament}\left[n(x)\Delta x\right]$, respectively, where the summation is along the $x$-axis (perpendicular to the page).
(b) Same as (a) but for another sample.
}\label{fig:filament+B+V}
\end{figure*}

\citet{2013ApJ...774L..31I} explained that the distorted field lines guide the gas flow into dense cores in a filamentary shape (see their Figure 1).
The created filaments are generally oriented in perpendicular direction to the magnetic field.
Filament formation is a common feature in a collision-compressed layer as shown in the other simulations of cloud-cloud collision \citep{2018PASJ...70S..53I}.
In this particular simulation by \citet{2013ApJ...774L..31I}, because the colliding two gas flows have different magnetic field orientations (perpendicular to each other), the directions of the resulting filaments are often crossing at 90 degrees.
This crossing feature disappears if the magnetic field orientations are similar between the colliding clouds.
Note that the crossing feature does not affect the mass distribution of the dense cores.
We look into more details in Figure \ref{fig:filament+B+V}, where the magnetic field lines and the velocity vector are projected in two typical regions with filaments.
In Figures \ref{fig:filament+B+V}(a)-1 and \ref{fig:filament+B+V}(b)-1 we find a common feature that the filaments are well aligned in a direction perpendicular to the field line.
The velocity vectors in Figures \ref{fig:filament+B+V}(a)-2 and \ref{fig:filament+B+V}(b)-2, on the other hand, do not show a systematic trend and are not always aligned with the field lines.
This suggests that the gas motion has both parallel and perpendicular components near the filaments, whereas the time average produces the net gas flow into the filaments.
We also notice that the line mass in a filament, which ranges from 20 to 150\,$M_{\solar}$\,pc$^{-1}$, tends to increase if the velocity vectors and field lines are in a similar direction as suggested in Figure \ref{fig:filament+B+V}(b).

\begin{figure*}
 \begin{center}
  \includegraphics{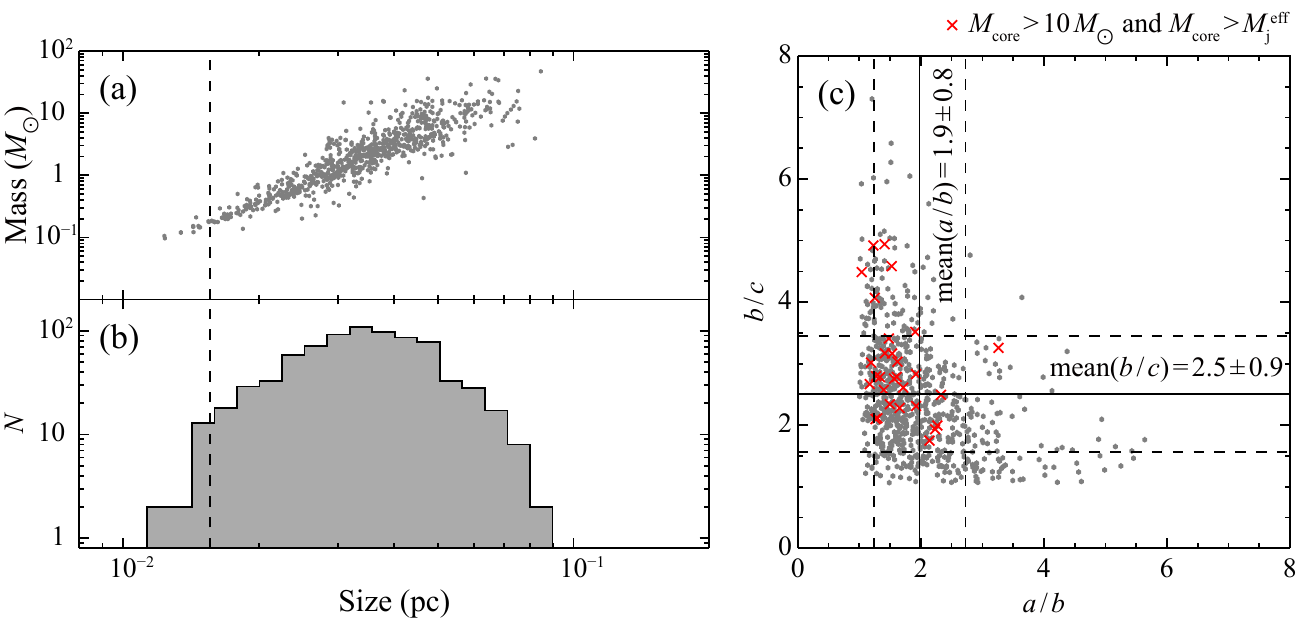} 
 \end{center}
\caption{(a) Mass-size diagram and (b) size histogram for the identified cores at $t=0.7$\,Myr.
The sizes are derived as geometric mean of the major-axis length $a$, second-major-axis length $b$ and minor-axis length $c$.
The vertical dashed lines in panels (a) and (b) show the pixel size ($1.6\times 10^{-2}$\,pc).
(c) Scatter plot between $b/c$ vs $a/b$.
The red crosses are the cores with $M_\mathrm{core}>10\,M_{\solar}$ and $M_\mathrm{core}>M_\mathrm{j}^\mathrm{eff}$.
The vertical solid- and dashed-lines show mean and standard deviation of $a/b$ and the horizontal ones show those of $b/c$.}\label{fig:size_diagrams}
\end{figure*}

\begin{figure}
 \begin{center}
  \includegraphics{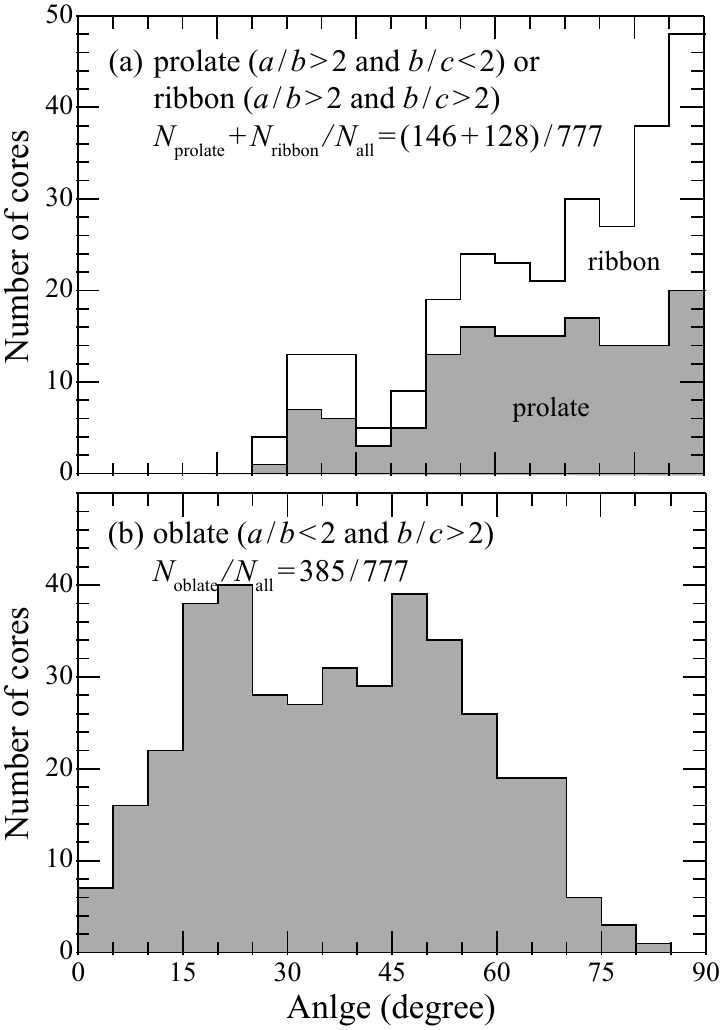} 
 \end{center}
\caption{(a) Histogram of angle between magnetic field vector and major-axis for prolate ($a/b>2$ and $b/c<2$, gray shaded) and ribbon-like morphology cores ($a/b>2$ and $b/c>2$).
(b) Histogram of angle between magnetic field vector and minor-axis for oblate cores with $a/b<2$ and $b/c>2$.
}\label{fig:angle_B_vs_core_axis}
\end{figure}

Each filament contains dense cores.
Formation of similar dense cores is observed in the non-MHD numerical simulations of a cloud-cloud collision by \citet{2014ApJ...792...63T}, which is due to cloud self-gravity.
In order to characterize the cores we applied the clump-find algorithm to the filaments at $n_\mathrm{threshold}$, and identified 777 cores in total at 0.7\,Myr as shown in Figure \ref{fig:filaments+cores}(b).
In each epoch the physical parameters including mass, size, average density of the dense cores are derived for density above $10^{5}$\,cm$^{-3}$.
The mass of a core is a sum of all of the pixels, and the size of a core is calculated as a geometrical mean of three major axes, which are derived by fitting a spheroid with three semi-major axes $a$, $b$, and $c$.
Figure \ref{fig:size_diagrams}(a) shows a scatter plot between mass and size of the dense cores, Figure \ref{fig:size_diagrams}(b) a size histogram, and Figure \ref{fig:size_diagrams}(c) a scatter plot between the axial ratios, $a/b$ and $b/c$.
Mass of a core is in a range 0.1 -- 100\,$M_{\solar}$ with size in a range of 0.01 -- 0.1\,pc.
The maximum mass of a core attained is close to 100\,$M_{\solar}$ with size of 0.1\,pc at density above $10^{5}$\,cm$^{-3}$, while it depends on the threshold density adopted.
The aspect ratios $a/b$ and $b/c$ are mostly in a range of 1 -- 3 with an average of 1.9 -- 2.5, indicating that they are not spherical: 15\% of the cores are spherical, 35\% prolate and 50\% oblate.
For the prolate and oblate cores we made histograms of an angle between the major axis and the field lines in Figure \ref{fig:angle_B_vs_core_axis}(a) and the minor axes and the field lines in Figure \ref{fig:angle_B_vs_core_axis}(b).
We find that the prolate cores are aligned within 40 degrees to the field lines, while the oblate cores have minor axes with an angle to the field lines in a broad range of 10 -- 70 degrees.
The former seems to be a natural result of the prolate core formation in a filament, while the oblate cores form more randomly.

\subsection{A core mass function}\label{subsec:CMF}
\begin{figure*}
 \begin{center}
  \includegraphics{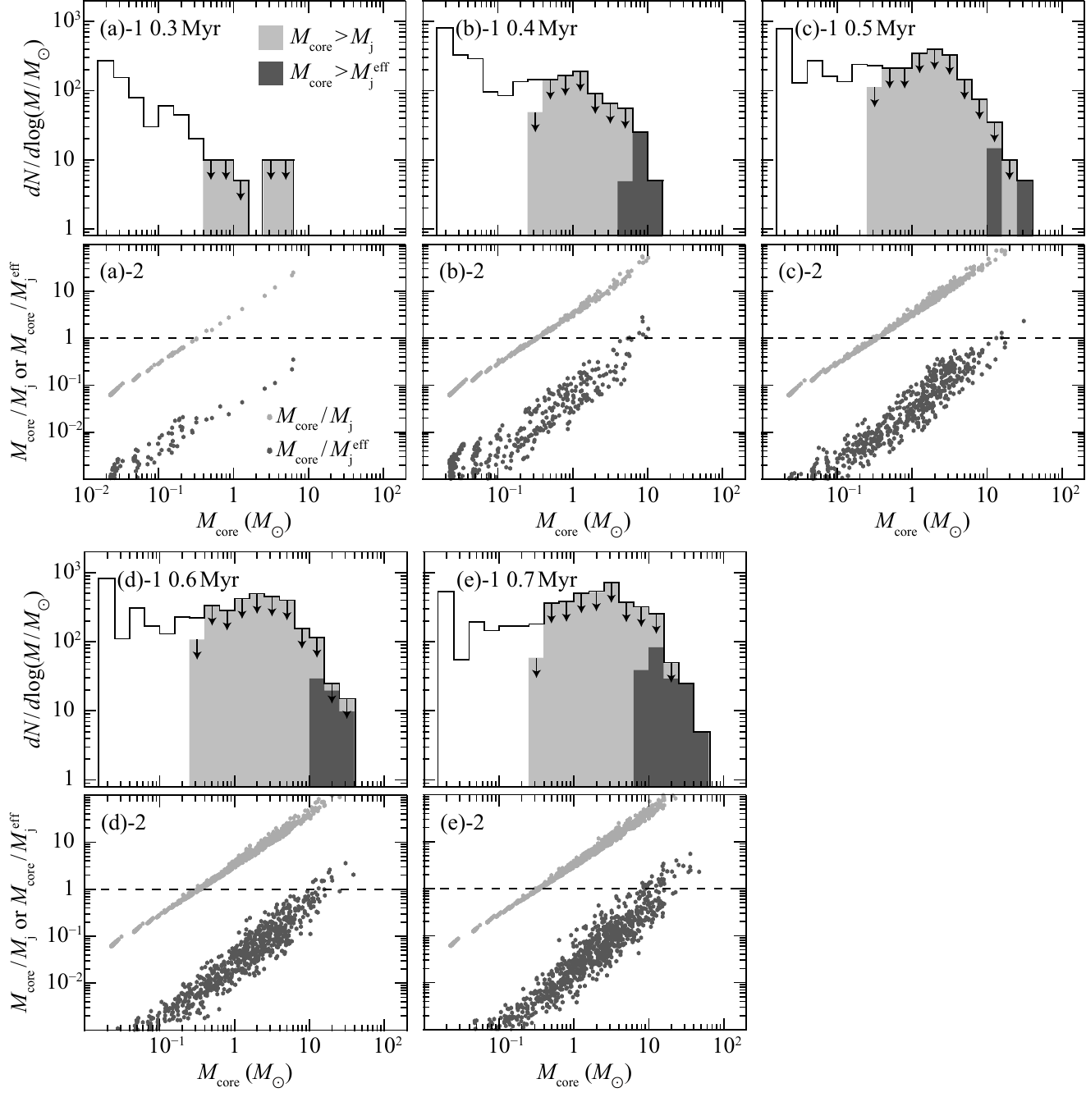} 
 \end{center}
\caption{\textit{upper panels:} Core mass function (CMF) for (a) $t=0.3$\,Myr, (b) 0.4\,Myr, (c) 0.5\,Myr, (d) 0.6\,Myr and (e) 0.7\,Myr.
Light- and dark-gray show CMF for $M_\mathrm{core}>M_\mathrm{j}$ and $M_\mathrm{core}>M_\mathrm{j}^\mathrm{eff}$, respectively.
\textit{lower panel:} $M_\mathrm{core}/M_\mathrm{j}$ and $M_\mathrm{core}/M_\mathrm{j}^\mathrm{eff}$ ratios plotted against $M_\mathrm{core}$.
The horizontal dashed lines show $M_\mathrm{core}=M_\mathrm{j}$ ($M_\mathrm{core}=M_\mathrm{j}^\mathrm{eff}$).
}\label{fig:CMF+EJM}
\end{figure*}

Figure \ref{fig:CMF+EJM} shows a core mass function, and the ratio of the core mass $M_\mathrm{core}$ to the effective jeans mass $M_\mathrm{j}^\mathrm{eff}$, which counts the turbulent and magnetic pressure in addition to the thermal pressure, and to the jeans mass $M_\mathrm{j}$, which counts only the thermal pressure, as a function of the core mass at five epochs.
The cores grow by the collisional compression.
We find an increase of the cores in a mass range 0.1 -- 10\,$M_{\solar}$ and the highest mass above $10^{5}$\,cm$^{-3}$ reaches $\sim 50$\,$M_{\solar}$ at 0.7\,Myr.
We see a clear signature of a top-heavy mass function extending beyond 10\,$M_{\solar}$ in Figure \ref{fig:CMF+EJM}.
The ratio between the core mass and the effective jeans mass shows that most of the cores with mass more than 10\,$M_{\solar}$ are gravitationally bound and will become high-mass protostars as shown by \citet{2013ApJ...774L..31I} and \citet{2018PASJ...70S..53I}.
Dense cores with mass less than 10\,$M_{\solar}$ are more weakly gravitationally bound.
They may form low-mass stars if the core mass is greater than the jeans mass depending on the details of the gas dynamics.
In this sense, the number density of the cores whose mass lies between 0.3\,$M_{\solar}$ and 10\,$M_{\solar}$ in Figure \ref{fig:CMF+EJM} gives an upper limit for the number density of the low-mass protostars.
It is unlikely that cores less massive than the jeans mass form stars, while the less massive cores may grow into a more massive ones by mass accretion and coagulation, depending on the time scale before dissipation by ionization due to the forming O stars.

The present model prediction of the core mass function consists of the high mass part for 6 -- 50\,$M_{\solar}$ (dark gray in Figure \ref{fig:CMF+EJM}(e)-1) and the low mass part for less than 20\,$M_{\solar}$ (light gray in Figure \ref{fig:CMF+EJM}(e)-1).
For the cores in the low mass part, it remains to be tested numerically if they really become gravitationally unstable to form stars.
The high mass part indicates the mass function of the gravitationally unstable cores and is robust.
The low mass part gives upper limits for the number of the cores leading to star formation.
If the cores can fragment further, the low mass part may be affected.
Given that the most unstable scale length $L$ of gravitational instability for a filament of width $W$ is $L = 4 W$ \citep{1992ApJ...388..392I}, the cores identified in our simulation having a typical aspect ratio of 1.9 -- 2.5 are the structures that are fragmented from a longer filament.
Further heavy fragmentation, therefore, does not seem to occur.
Recent simulations that followed the gravitational collapse phase of a filamentary core created by a cloud collision show that the massive filamentary core collapses into a massive sink \citep{2018PASJ...70S..53I}.
Note that we never claim that each massive core found in our analysis certainly evolves into one massive star with high star formation efficiency.
The cores can fragment during collapse phase.
However, recent theoretical studies suggest that once massive compact core is created, the collapse of which will eventually create massive star(s), i.e., collapse of a massive core would not result in formation of many low-mass stars due to heavy fragmentations 
\citep{2009Sci...323..754K,2016MNRAS.463.2553R}
In addition, the massive star formation time scale is very short in the order of $5\times 10^{4}$\,yr once the core is formed \citep{2009Sci...323..754K}, in the order of the free fall time $10^{5}$\,yr for $10^{5}$\,cm$^{-3}$.
It is therefore likely that the star(s) begin to warm up the surrounding gas rapidly, and further fragmentation of the rest of the cores is not very effective \citep{2006ApJ...641L..45K}.
All these features allow us to compare our core analysis with observations at least in high-mass star forming regions.

\section{Properties of the dense cores and O stars}\label{sec:properties_cores}
The present results on dense cores provide an insight into the O star formation by cloud-cloud collision.
We first describe the formation of individual O stars in dense cores, and, second, present the predicted mass function and spatial distribution of multiple O stars. 

\subsection{O star formation and subsequent evolution in the shock compressed layer}
In the shock-compressed layer, dense cores smaller than 0.1\,pc with maximum mass close to 100\,$M_{\solar}$ are created.
These cores have density more than $10^{5}$\,cm$^{-3}$ and become gravitationally unstable.
\citet{2013ApJ...774L..31I} showed that the mass accretion rate in the shock-compressed layer is greater than $\sim 10^{-4}$\,$M_{\solar}$\,yr$^{-1}$ due to the enhanced effective sound speed, which is large enough to overcome the radiation pressure to form high-mass stars \citep{1987ApJ...319..850W}.
The formation time scale of a 15\,$M_{\solar}$ O star is estimated to be $15\,M_{\solar} /10^{-4}\,M_{\solar}\,\mbox{yr}^{-1} = 0.15\,\mbox{Myrs}$ if a constant mass accretion rate is assumed.
In this mass accretion phase, we expect molecular outflow is generated as driven by the accretion disk in a dynamical timescale of $\sim 10^{4}$\,yr (e.g., \cite{2008ApJ...676.1088M}).
When the star becomes as massive as 15\,$M_{\solar}$, its surface temperature reaches 30000\,K \citep{2000asqu.book.....C}, which is high enough to emit significant ionizing photons to its surroundings.
The ionization then proceeds at a typical speed of more than a few km\,s$^{-1}$ (e.g., \cite{2005ApJ...623..917H}), and in a few times 0.1\,Myrs a radius of $\sim1$\,pc will be ionized, while a significant portion of the neutral gas generally remains unionized outside the radius.

The O star formation is a natural outcome of cloud-cloud collision, a rapid process which converts the low-density molecular gas to dense gas in 0.1\,Myrs.
It is essential that the compression is done not by self-gravity but by the magnetic-field guided supersonic flow.
This rapidness is the key to collect large mass of 100\,$M_{\solar}$ into a 0.1\,pc volume.
If the collect process is slow in the order of the free fall time ($\sim$\,Myr for density 1000\,cm$^{-3}$), it is likely that the gas is consumed by formation of low-mass stars before being collected in the volume.

Gravitational contraction of a dense core is numerically simulated by \citet{2009Sci...323..754K} at highest resolution of 100\,AU, who used 100\,$M_{\solar}$ within a radius of 0.1\,pc having a power law distribution at the initial condition.
The results show that two stars with mass of 40\,$M_{\solar}$ and 30\,$M_{\solar}$ with remaining 30\,$M_{\solar}$ gas, where star formation efficiency (SFE) is 70\%.
The two stars are formed rapidly in a timescale of $5\times 10^{4}$\,yr.
Their core is similar to the typical massive unstable cores formed under cloud-cloud collision of the present work.
These authors adopted the typical massive core observed in high mass star forming regions, and did not explore the core formation mechanism.

\subsection{Cluster formation in the collision-compressed layer}

\subsubsection{Formation of filaments and O stars in the compressed layer}
\begin{figure*}
 \begin{center}
  \includegraphics{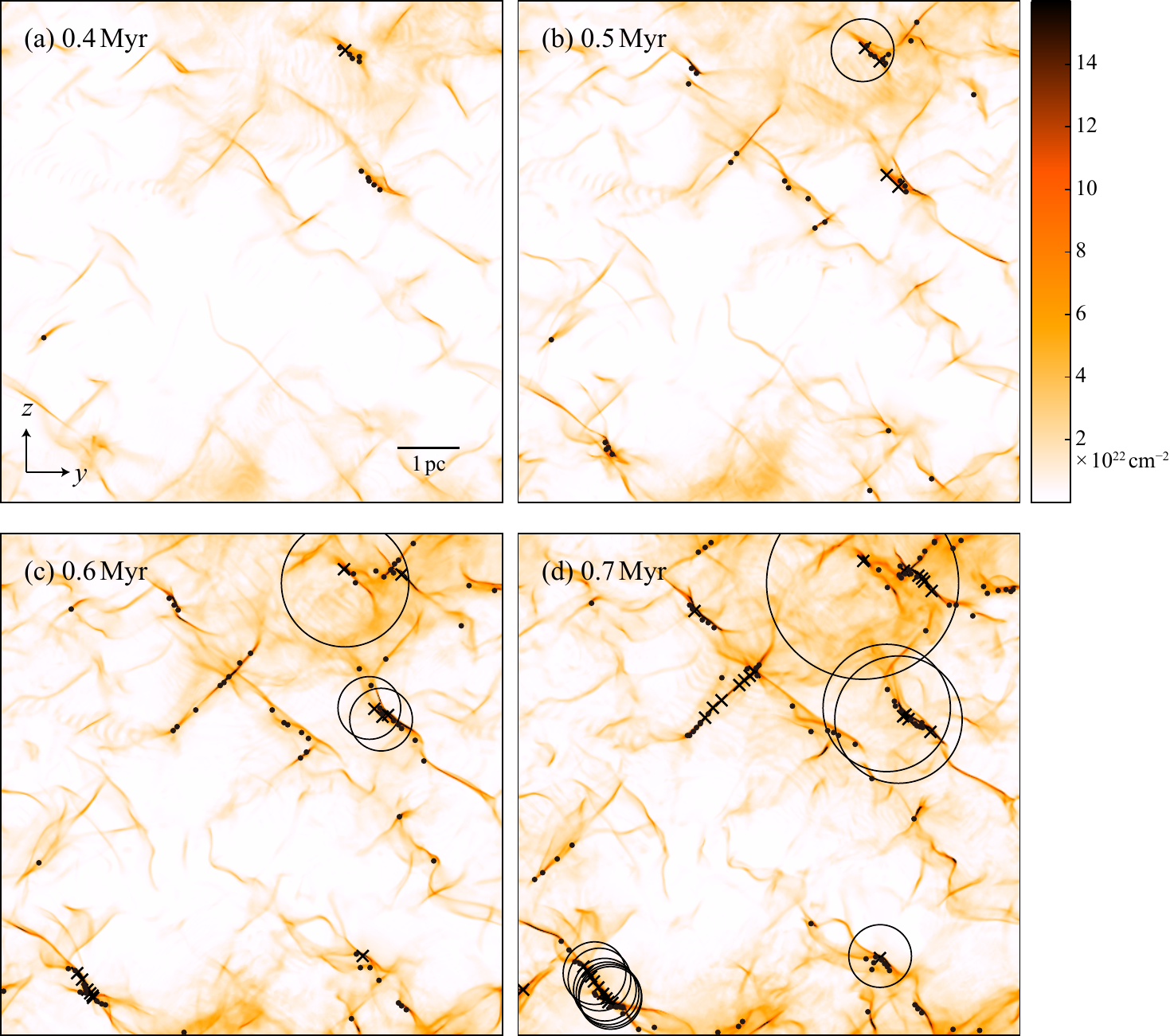} 
 \end{center}
\caption{
Column density map of H$_{2}$ (low density gas with $< 10^{4}$\,cm$^{-3}$ is excluded) in the $y$-$z$ plane at (a) $t=0.4$\,Myr, (b) 0.5\,Myr, (c) 0.6\,Myr and (d) 0.7\,Myr.
The crosses show the positions of massive cores with $M_\mathrm{core}>10\,M_{\solar}$ and $M_\mathrm{core}>10\,M_\mathrm{j}^\mathrm{eff}$, and the dots show intermediate mass cores with $M_\mathrm{core}=5 \mbox{ -- } 10\,M_{\solar}$.
Circles show ionization front in ``HII regions'' which proceeds at a uniform speed of $X$\,km\,s$^{-1}$ after dense cores with $M_\mathrm{core}>10$\,$M_{\solar}$ and $M_\mathrm{core}>10\,M_\mathrm{j}^\mathrm{eff}$ appear.
}\label{fig:Ostars+HIIregions}
\end{figure*}

A cloud-cloud collision produces the filamentary morphology of the compressed layer and the dense-core distribution confined to the filaments as shown in Figure \ref{fig:Ostars+HIIregions}.
The O stars formed will not move by more than 0.2\,pc within 0.1\,Myrs, the typical timescale of the O star formation, for an average velocity of dense cores of $\sim 1$\,km\,s$^{-1}$ (Figure \ref{fig:velocity}).
This indicates that the O star distribution is determined by the collision-formed filamentary distribution until gravitational relaxation becomes effective later, in more than Myr.
The evolution of the stellar system in this phase is essentially similar to that given by $N$-body simulations (e.g., \cite{2015MNRAS.449..726F}).
The filaments will be destroyed in the order of 0.1\,Myrs by the ionization once the stars become as massive as 15\,$M_{\solar}$, and the final mass of the formed stars, i.e., the O stars and their neighbors, are determined by the mass of the dense cores and the duration of the mass accretion until full ionization.
In the following we assume for simplicity that the mass of the gravitationally unstable dense cores is converted into the stellar mass, while the conversion efficiency may be somewhat less than 1.0.
The numerical simulation result for instance shows that SFE is 0.7 for a 100\,$M_{\solar}$ core within 0.1 pc \citep{2009Sci...323..754K}.
In the present simulations no ionization is incorporated.
In Figure \ref{fig:Ostars+HIIregions} we draw a circle of the ionization front for each ``O star'' by assuming for simplicity that the ionization front proceeds at a uniform speed of 3\,km\,s$^{-1}$ after a dense core of 10\,$M_{\solar}$ having density $10^{5}$\,cm$^{-3}$ appears.
This allows us to have a rough idea on the dispersal of the filaments by ionization.

\begin{figure*}
 \begin{center}
  \includegraphics{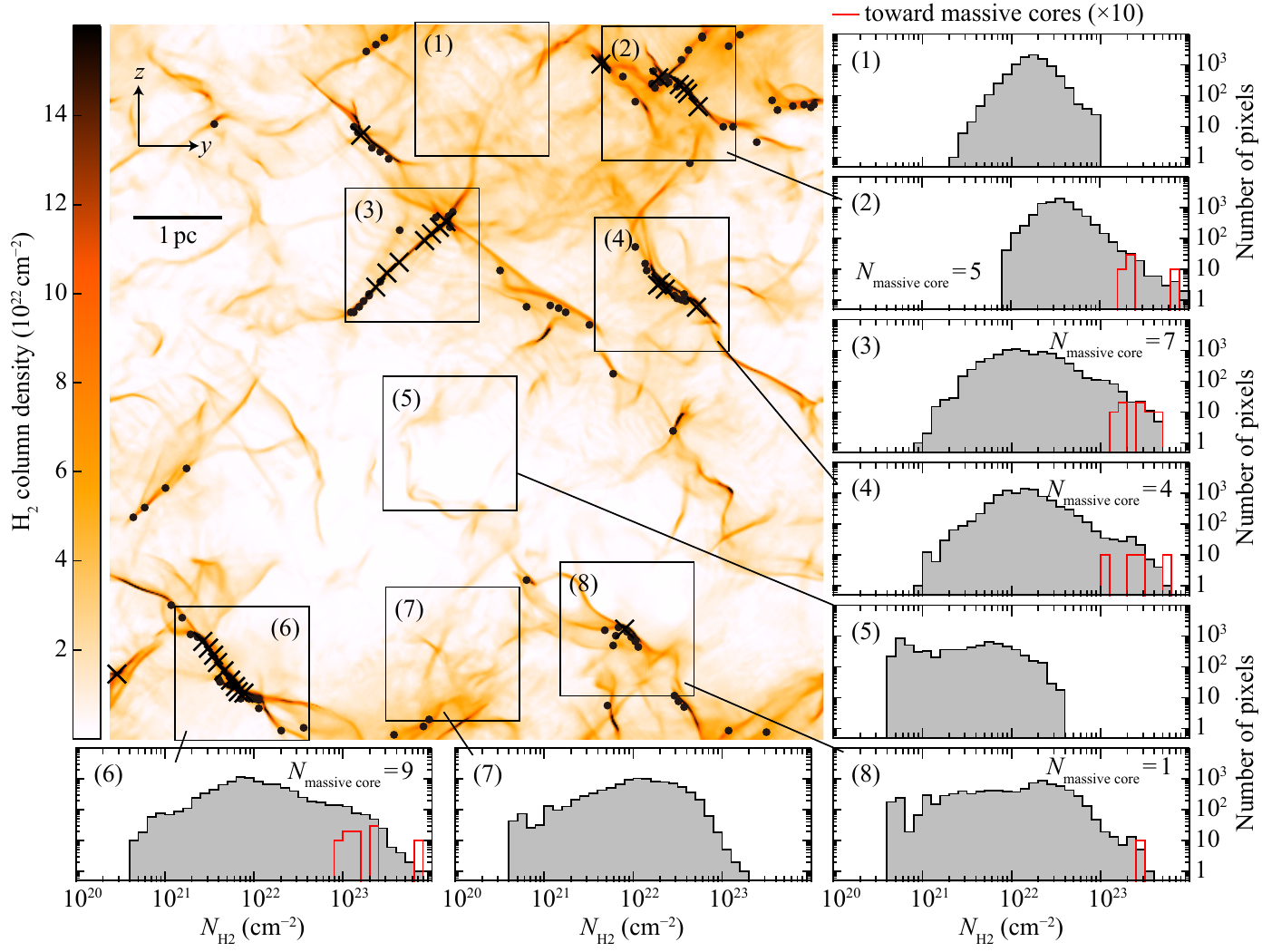} 
 \end{center}
\caption{
Column density map of H$_{2}$ (low density gas with $< 10^{4}$\,cm$^{-3}$ is excluded) in the $y$-$z$ plane at $t=0.7$\,Myr  (top-left panel) and column density histogram in the 8 regions of $1.5\,\mbox{pc} \times 1.5\,\mbox{pc}$ (panels (1) -- (8)).
The crosses in the top-left panel show the positions of massive cores with $M_\mathrm{core}>10\,M_{\solar}$ and $M_\mathrm{core}>10\,M_\mathrm{j}^\mathrm{eff}$, and the dots show intermediate mass cores with $M_\mathrm{core}=5 \mbox{ -- } 10\,M_{\solar}$.
}\label{fig:coldens_and_CMF}
\end{figure*}

We obtain a further insight into the O star formation by inspecting column density in the filaments.
Figure \ref{fig:coldens_and_CMF} shows the column density distribution in space for 8 regions of $1.5\,\mbox{pc} \times 1.5\,\mbox{pc}$ in the $y$-$z$ plane selected by the number of O stars and the distribution function of column density in each box.
Figure \ref{fig:coldens_and_CMF} indicates that O stars can be formed in isolation and/or in a cluster depending on the resultant column density.
In the distribution function of column density in Figure \ref{fig:angle_B_vs_core_axis}, the typical column density peaks are found at around $10^{22}$\,cm$^{-2}$.
Formation of nearly ten O stars occurs in a filament with high column density tails beyond $10^{23}$\,cm$^{-2}$ (Regions (2), (3), (4), and (6)).
On the other hand, in a region of a single or no O star formation column density distribution is not extended beyond $10^{23}$\,cm$^{-2}$. 

\subsubsection{Cluster formation scenario}
A cloud-cloud collision creates massive dense cores, precursors of an O star, which are clustered in a filament. In order to illuminate the whole process of a cluster formation, we present a formation scenario of an O-star cluster as a two-step process, i.e., (i) pre-collision phase, and (ii) post-collision phase, in the following.
Similar discussion is given for two individual objects RCW38 and the Orion Nebula Cluster (ONC), from an observational point of view \citep{2016ApJ...820...26F,2018ApJ...859..166F}.

In the pre-collision phase, the two clouds independently form only low-mass stars from low-mass cores without mutual interaction if they are dense enough: a typical observed case is the Taurus cloud complex forming $\sim 20$ low mass stars in a cloud having mass of $\sim 1000\,M_{\solar}$, size of $\sim 5$\,pc and column density $(1 \mbox{--} 10) \times 10^{21}$\,cm$^{-2}$ \citep{1995ApJ...445L.161M,1996ApJ...465..815O,1998ApJ...502..296O,2002ApJ...575..950O}.
The timescale of this phase can be as long as several Myrs, forming low-mass stars at a mass accretion rate of $10^{-6}$\,$M_{\solar}$\,yr$^{-1}$ in the non-shocked condition as long as the cloud dispersal by outflows/winds is not substantial. 
The core mass function is shown in \citet{2002ApJ...575..950O}, which indicates that the dense core of density $10^{5}$\,cm$^{-3}$ have mass range from 1 to 10\,$M_{\solar}$, which can be roughly converted to a stellar mass range from 0.1 to 1\,$M_{\solar}$.

Suppose an accidental collision between such two clouds happens at a supersonic velocity in the order of 10\,km\,s$^{-1}$, the collision rapidly creates the collision-compressed layer in $\sim 0.1$\,Myrs to form massive dense cores  (the post-collision phase).
The moment of the collision is subject to a stochastic process among the clouds in the Galactic disk with a typical mean free time of less than 10\,Myrs depending on the cloud number density as shown by numerical simulations (e.g., \cite{2018PASJ...70S..59K}; see also \cite{2014MNRAS.445L..65F,2014prpl.conf....3D}).
The number of O star(s) depends on the cloud column density.
Based on the previous observations of cloud-cloud collisions, \citet{2018ApJ...859..166F} showed that a single O star can be formed for total column density of $10^{22}$\,cm$^{-2}$, and that formation of nearly ten O stars requires total column density as high as $10^{23}$\,cm$^{-2}$ at a 1-pc scale (see also \cite{2019PASJ..tmp..127E}).
Clouds with column density below $10^{22}$\,cm$^{-2}$ will grow in mass by collision without O star formation.

Subsequently, the massive dense cores form O stars which ionize the clouds and suppress subsequent star formation, details of which depends on the number of the O stars, and density and size of the clouds.
In the end of this phase a cluster becomes exposed by ionization, and the emerging cluster consists of old low-mass stars extended over the clouds and young O star(s) localized within the colliding area.
Consequently, the age of the low-mass members can be distributed from 1\,Myr to several Myrs as determined by the collision mean free time, and the age of the O stars is peaked just after the collision with a short duration in the order of 0.1\,Myrs.
Such a short time duration is supported by the detailed observations of stellar ages in two clusters NGC3603 and Westerlund1 \citep{2012ApJ...750L..44K}.
In $\sim 1$\,Myr after the O star formation the ambient gas will be fully ionized and dispersed as seen in NGC3603 and Westerlund1.
After the ionization, gravitational relaxation will redistribute the filamentary O stars into a more centrally concentrated stellar cluster in a few Myr, as is modeled by the $N$-body numerical simulations (e.g., \cite{2015MNRAS.449..726F}).

For example, the ONC and RCW38 correspond to the earliest post-collision phase of 0.1\,Myr after collision, and NGC3603 is of 2\,Myrs after the ionization. 
The pre-collision phase is not easily recognized observationally at a usual distance of O stars more than 1\,kpc because a pre-collision cluster without O stars is not obvious in the Galactic plane due to extinction and contamination. 

\section{Comparison with observations of O stars triggered by a cloud-cloud collision}\label{sec:comparison_with_obs}
Recent studies show that cloud-cloud collision is an essential process in high-mass star formation.
Observationally, cloud-cloud collision is found in many high-mass stars associated with molecular clouds which are not fully ionized.
Here we compare the present theoretical results with observations of cloud-cloud collisions in the youngest phase of $\sim 0.1$\,Myrs, where we are able to find the O star distribution just after their formation without ionization or gravitational relaxation.
The key features of the O star formation derived in the present work and \citet{2013ApJ...774L..31I} are summarized as follows;

a. Cloud-cloud collision is a mechanism of rapid formation of massive dense cores with density of $10^{5}$\,cm$^{-3}$ up to $\sim 50$\,$M_{\solar}$ in 0.7\,Myr, which is characterized by a top-heavy mass function.
The massive dense cores are gravitationally unstable and lead to O star formation under the high-mass accretion rate.
Cloud-cloud collision is therefore a mechanism of O star formation.
Less massive dense cores, some of which may be transient, are also formed more in number than the massive dense cores in the collision compressed layer, while a small fraction of them perhaps becomes gravitationally unstable to form low-mass stars.

b. The colliding gas flow of 10\,km\,s$^{-1}$ is quickly decelerated to around 1.5\,km\,s$^{-1}$ in the collisional-shocked layer, where the velocity vector becomes randomized.
The collision-compressed layer of $\sim 1$\,pc thickness becomes filamentary with 0.1-pc width, where the gas flow is guided to a filament by the magnetic field lines.
The formation of the filament is not due to the self-gravity and is, therefore, more rapid in $\sim 0.1$\,Myr than the free fall time, $\sim 1$\,Myr for 1000\,cm$^{-3}$, which is essential in O star formation.
Once a filament is formed, the filament grows in mass by the incoming flow.
In the filament the massive dense cores are formed by self-gravity in a 0.1\,pc scale in free-fall time of 0.1\,Myrs for density above $10^{5}$\,cm$^{-3}$.

c. The massive dense cores form O stars which are confined in the filament, and the O star distribution is filamentary, having the same velocity field with the collisional shocked layer.
When a mature O star of more than 15\,$M_{\solar}$ forms, the ionization by the O star becomes effective to disperse the gas and quenches star formation within a few pc of the O star(s).
The ionization removes gas and the O star distribution evolves by gravitational relaxation in a Myr timescale.

For a comparison with the observations we use the two properties of the dense cores of the present work, the separation between the cores and the core mass function.

\begin{figure*}
 \begin{center}
  \includegraphics{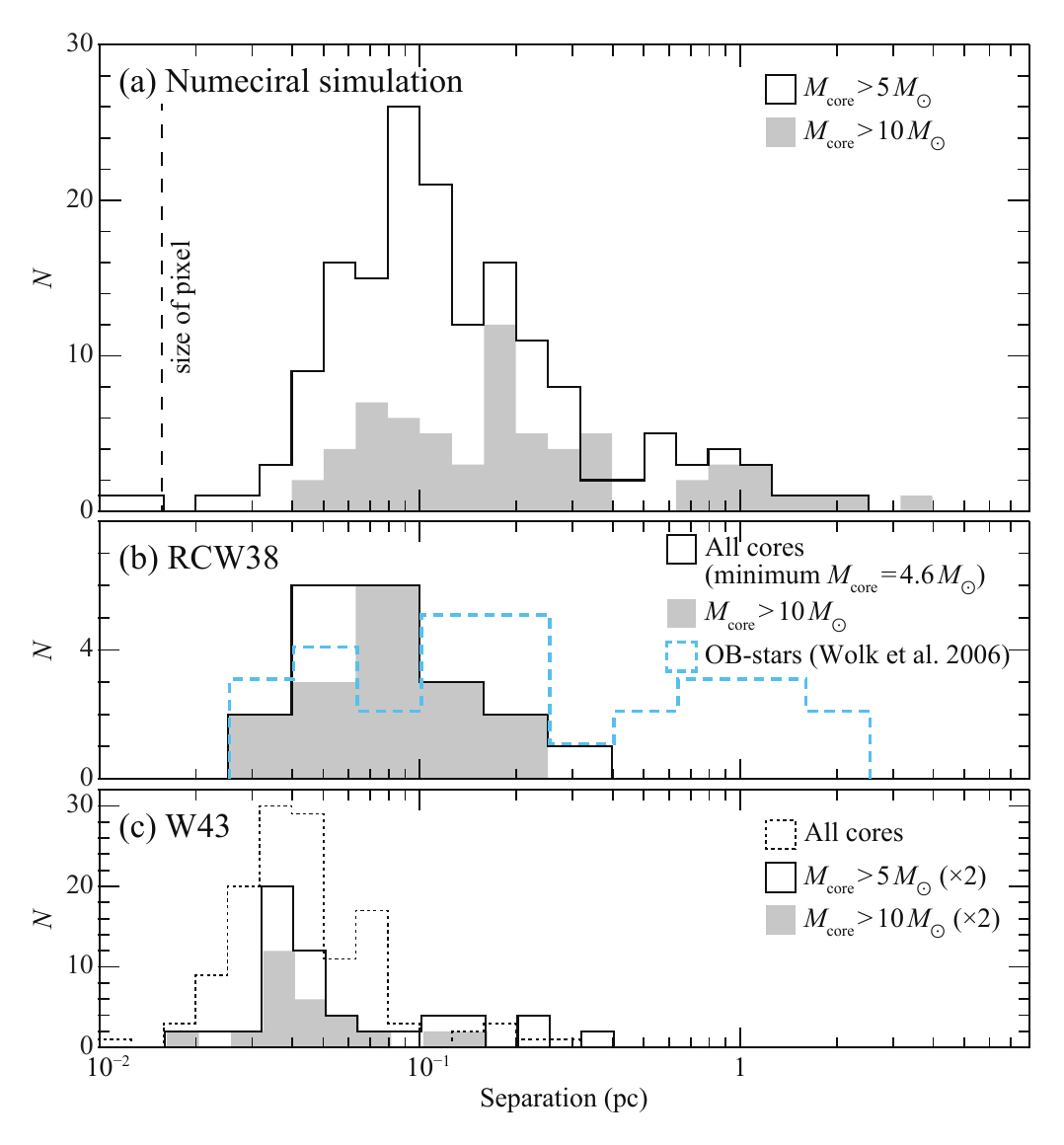} 
 \end{center}
\caption{(a) Histogram of core-to-core separation for $M_\mathrm{core}>5\,M_{\solar}$ at $t=0.7$\,Myr and those for $M_\mathrm{core}>10\,M_{\solar}$ (gray shaded).
Here, the separations are given as edge-lengths of two-dimensional (projected along the $x$-axis) minimum-spanning-trees (MSTs) of cores.
(b) Same as (a) but for RCW38 dense condensations \citep{2019arXiv190707358T} and OB-star candidates \citep{2006AJ....132.1100W}.
(c) Same as (a) but for W43 cores \citep{2018NatAs...2..478M}.
}\label{fig:separation}
\end{figure*}

\begin{figure*}
 \begin{center}
  \includegraphics{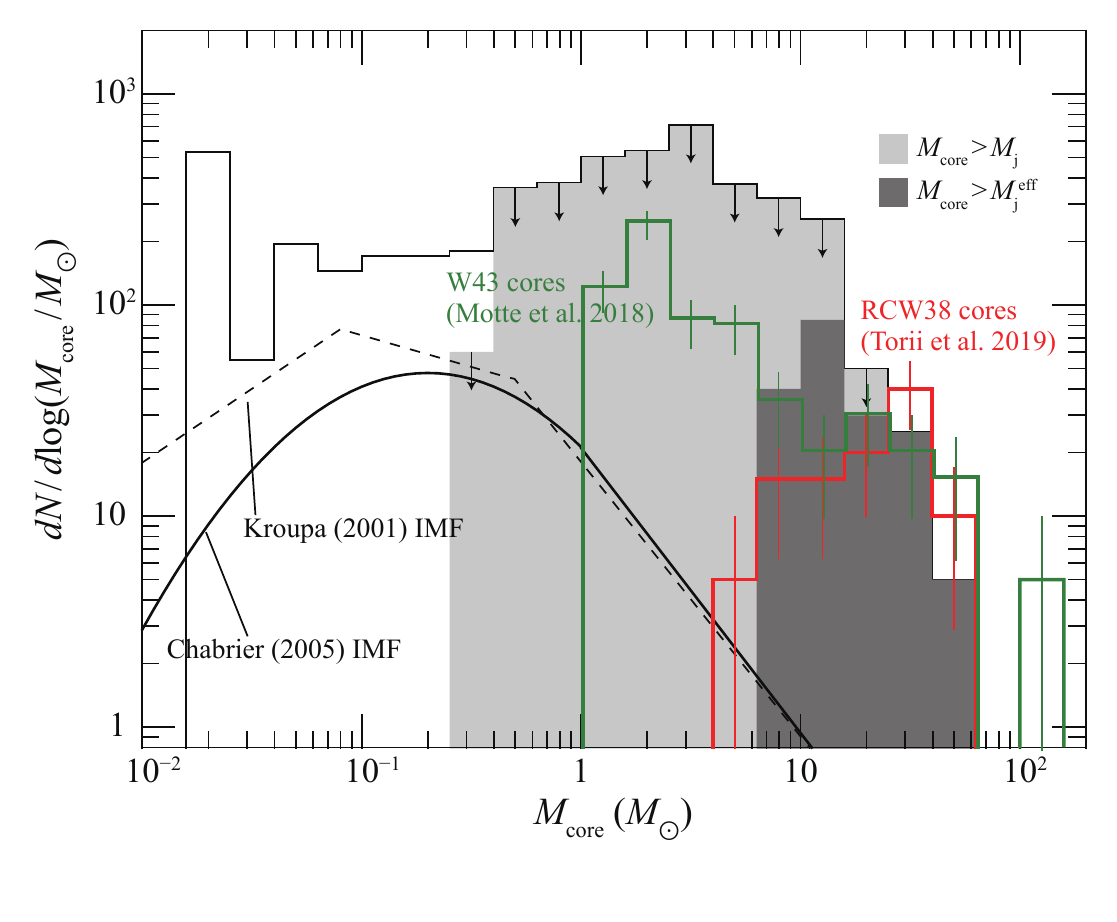} 
 \end{center}
\caption{
CMF at $t=0.7$\,Myr (identical to Figure \ref{fig:CMF+EJM}(e)-1).
Those for RCW38 cores \citep{2019arXiv190707358T} and W43 cores \citep{2018NatAs...2..478M} are superimposed.
The error bars correspond to $\sqrt N$ statistical uncertainties.
The dashed line shows the single-star IMF of  \citet{2001MNRAS.322..231K} and the solid curve shows the system IMF by \citet{2005ASSL..327...41C}.
}\label{fig:CMF}
\end{figure*}

Figure \ref{fig:separation}(a) shows a histogram of the separation between dense cores with mass more than 5\,$M_{\solar}$ in the $y$-$z$ plane.
The separation is peaked at 0.1\,pc and is extended up to a few pc.
The cores more massive than 10\,$M_{\solar}$ show two peaks at $\sim 0.1$\,pc and $\sim 1$\,pc, where the small value reflects the core distribution within a filament and the large one the core distribution between filaments.
The distribution provides a test tool for a comparison with observations. 
Figure \ref{fig:CMF} shows another comparison of the core mass function in the collision compressed layer taken from Figure \ref{fig:CMF+EJM}.
The core mass function in cloud-cloud collision is top-heavy with a negligible contribution of dense cores less than 1\,$M_{\solar}$.
The initial mass function  (IMF) of field stars which is shifted to lower mass by more than an order of magnitude as compared with that in cloud-cloud collision (e.g., \cite{2001MNRAS.322..231K,2005ASSL..327...41C}), whereas a direct comparison between cores and stars requires exact SFE and multiplicity of stars.
The separation between the cores are compared both with the observed cores and stars in the following by assuming that they do not move over a distance of more than 0.1 pc in 0.1 Myr.
The cores in the core mass function include those which are not gravitationally unstable.
So, the core mass function is not the stellar mass function, and a core may form two or more stars with SFE lower than 1.0.
Such  a comparison is valid as long as it is made only for the cores but not for the stars.

Super star clusters in the Milky Way include 10 -- 20 O stars in a pc scale in addition to $\sim 10^{4}$ low mass stars more extended (see for a review \cite{2010ARA&A..48..431P,2018ASSL..424....1A}).
Among them the very rich clusters including O stars as young as 0.1\,Myr are the primary site to test the collision signatures, separations between dense cores (or O stars) and a core mass function, which will be rapidly lost by ionization in Myr.
 
\subsection{RCW38}\label{ssec:RCW38}
As the best cluster for comparison we focus on RCW38 the youngest super star cluster where O star formation was triggered by a cloud-cloud collision according to \citet{2016ApJ...820...26F}.
Most recently, \citet{2019arXiv190707358T} mapped the CO clouds with ALMA at high resolution of $\sim 2^{\prime\prime}$ in C$^{18}$O $J=2 \mbox{--} 1$ emission and resolved 21 dense cores larger than 0.01\,pc with a mass detection limit of $\sim 6$\,$M_{\solar}$.
The O star candidates \citep{2006AJ....132.1100W,2011ApJ...743..166W} are distributed in filamentary distribution of $\sim 1$\,pc length toward the collisional area of a 0.5-pc radius as shown in Figure 1 of  \citet{2019arXiv190707358T}.
Figure \ref{fig:separation}(d) shows that separations between the dense cores/O stars  are consistent with the present results.
In Figure \ref{fig:CMF} the core mass function is compared, showing also good correspondence in a virial mass range of 8 -- 50\,$M_{\solar}$.
The C$^{18}$O cores in RCW38 are often gravitationally unstable as shown by the virial mass which is calculated to be smaller than the molecular mass (see their Table 3).
They are probable ``protostellar'' cores, which is also supported by two of the RCW38 cores with protostellar outflow.
Higher sensitivity observations in RCW38 will reveal lower mass cores which cover a wider mass range.
The top-heavy core mass function in RCW38 is consistent with the collisional compression, and also the separations of the massive dense cores and the O stars.
The dense gas which is included in the detected dense cores is 360\,$M_{\solar}$, which  corresponds to 33\% of the total mass of the dense gas detected in the C$^{18}$O ($J=2\mbox{--}1$) emission \citep{2019arXiv190707358T}, although cores with mass below 6\,$M_{\solar}$ are not spatially resolved.
This fraction may be compared with the total mass of the unstable massive cores in the present simulations, 650\,$M_{\solar}$, which corresponds to 22\% of the total mass of the gas with density above $10^{5}$\,cm$^{-3}$.
The both show a significantly high fraction of the massive cores.

We test the physical parameters of the clouds into more detail.
The total number of the massive dense cores and O star candidates in RCW38 is $\sim 40$ for an area of $\sim 0.5\,\mbox{pc}\times 0.5\,\mbox{pc}$ and the projected density of cores/O stars is $\sim 160$\,pc$^{-2}$, ten times higher than in Region (6) of the present results.
The age of RCW38 is in the order of 0.1\,Myrs, which is significantly shorter than 0.7\,Myrs of the present model.
These suggest that the initial density of the parent cloud of RCW38 may be significantly higher than the present initial density 300\,cm$^{-3}$.
\citet{2013ApJ...774L..31I} showed a case of the initial density 1000\,cm$^{-3}$, three times higher than the present model (Model No.\ 4 in \cite{2013ApJ...774L..31I}) which indicates more rapid dense core formation and higher column density in the compressed layer.
A rough estimate indicates that the initial density more than $10^{4}$\,cm$^{-3}$ may be appropriate to explain these differences, which needs to be confirmed by further MHD simulations.
It is desirable to have more objects to be tested, whereas most of the other massive clusters in the Milky Way studied so far into detail are not so young as RCW38.
The only possible exception is the ONC, which is worth testing by a uniform survey for dense cores with ALMA over the collisional area.
The collisional age of the O stars in the ONC is 0.1\,Myrs \citep{2018ApJ...859..166F} and such observations will shed new light on the issue.

\subsection{W43}\label{ssec:W43}
W43 is a remarkable mini-starburst in the Milky Way which attracted keen interest in understanding high mass star formation.
Recently, \citet{2020arXiv200110693K} made a detailed analysis of the molecular clouds in W43 and found evidence for multiple cloud-cloud collisions in triggering high-mass star formation.
In one of the proto clusters in W43 Main, new ALMA observations were used to resolve dense cores with unprecedented details by \citet{2018NatAs...2..478M}.
These authors obtained a core mass function of 130 dense cores with density $\sim 10^{7}$ -- $10^{10}$\,cm$^{-3}$ covering a mass range 1 -- 100\,$M_{\solar}$ in dust continuum emission with a mass detection limit of 1.6\,$M_{\solar}$, while linewidths and other kinematic information are not available.
The cores may be protostellar, if we consider the active high mass star formation in W43 Main \citep{2018NatAs...2..478M,2020arXiv200110693K}, while their gravitationally stability is not shown by spectroscopic data.
Figure 1 of \citet{2018NatAs...2..478M} shows the dust image  where we find highly filamentary dust distribution with dense cores.
The filaments have $\sim 0.1$-pc width and $\sim 0.5$-pc length and the mass of the dense cores are derived.
The formation of filaments by a cloud-cloud collision is a plausible mechanism applicable to W43.
The separations between the dense cores as shown in Figure \ref{fig:separation}(e) have a peak at $\sim 0.4$\,pc, which is similar but somewhat smaller than in RCW38.
The core mass function is plotted from \citet{2018NatAs...2..478M} in Figure \ref{fig:CMF} and shows that the core mass function is consistent with the present one in particular at the high-mass end of the present function in a range 8 -- 50\,$M_{\solar}$.
The massive dust cores more than 6\,$M_{\solar}$, 620\,$M_{\solar}$, occupy 70\% of the total mass of the detected dust cores, $880\,M_{\solar}$.
This may indicate a more top-heavy mass function than in RCW38, and the higher density environment of W43 may be responsible for it.

\section{Dependence on the model parameters}\label{sec:model_dependency}
The present results depend on the parameters adopted in the simulations by \citet{2013ApJ...774L..31I}.
The model is idealized as compared with the real cloud-cloud collision, whereas no fine tuning of the parameters is made.
We discuss possible adjustments of the parameters for a more sophisticated model.

The size of the present collision area $8\,\mbox{pc} \times 8\,\mbox{pc}$ is larger than the typical area of 1\,pc$^{2}$ in the cloud-cloud collisions in the Milky Way.
This difference perhaps causes higher gas peak column density in the present results than a smaller area by collecting mass in directions perpendicular to the collision.
The size of the model cloud length, which is assumed to be as long as 7\,pc by setting the final epoch to be 0.7\,Myr, is also longer than the typical real clouds.
The small length can lead to a decrease of collision density of one of the clouds, and thus needs a non-steady simulation where one of the flows decreases in density in time.

The other effect which can be significant is the stellar feedback, in particular, ionization which is not incorporated in the present model.
Such inclusion of ionization was undertaken by \citet{2018PASJ...70S..54S} in a two cloud collision case, and will be made separately from the present paper as a future work.
This will allow us to have a more realistic picture of the formed cluster.  

Another concern is density and velocity adopted as the initial condition.
The active star formation in RCW38 and W43 may suggest higher initial density than the present model as shown by the shorter timescale less than 0.1\,Myr of O star formation in RCW38 and the numerous O star candidates in W43.
We will need therefore test density dependence of the star formation.
Such a wider range of the model parameters will help to gain an insight into the outcomes  in cloud-cloud collision.

Finally, the relationship between the dense cores and the stellar mass is to be better quantified, while in the present work we simply assume that the core mass at density higher than $10^{5}$\,cm$^{-3}$ roughly corresponds to the stellar mass.
It is probable that the conversion is incomplete and a fraction of the core mass becomes a star.
Such an attempt is found in \citet{2018NatAs...2..478M}.
We will explore this issue along with the above tasks in future modeling.

\section{Conclusions}\label{sec:conclusions}
In order to obtain a better insight into the star formation triggered by cloud-cloud collision, we analyzed the simulation data of colliding molecular flows made by \citet{2013ApJ...774L..31I}.
The main conclusions of the present study are summarized below.

\begin{enumerate}
\item The two supersonic molecular flows in head-on collision at a relative velocity of 20\,km\,s$^{-1}$ create a shock compressed layer of 1-pc thickness. In the layer, velocity is decelerated rapidly and gas density becomes higher by two orders of magnitude.
The gas distribution is characterized by filamentary distribution at pc scale, which is elongated perpendicular to the field direction.
The gas flow along the field increases the filamentary mass, and dense cores are formed in the filaments.
The dense cores have size and mass of 0.03 -- 0.08\,pc and 8 -- 50\,$M_{\solar}$, respectively.
The most massive cores reach $\sim 50$\,$M_{\solar}$ at density above $10^{5}$\,cm$^{-3}$ within a radius of 0.1\,pc and the cores more massive than 10\,$M_{\solar}$ become gravitationally unstable.
They are plausible precursors of high-mass protostars.
\item The present results provide a comprehensive scenario as an alternative to the two conventional scenarios of high-mass star formation, i.e., the competitive accretion and the monolithic collapse.
Cloud-cloud collision realizes high-mass accretion rate of $10^{-4}$ -- $10^{-3}$\,$M_{\solar}\,\mbox{yr}^{-1}$ and offers a condition suitable to high-mass star formation.
 It is shown that cloud-cloud collision provides a versatile scenario which accommodates various O star distribution for a wide mass range of the parent clouds.
Observations show that single O star formation is possible for merged column density of $\sim 10^{22}$ \,cm$^{-2}$ and cloud mass of around 100\,$M_{\solar}$.
It is also possible that tens of O stars are formed for column density distribution extending above $10^{23}$\,cm$^{-2}$ even for small colliding cloud mass less than $10^{3}$\,$M_{\solar}$.
This offers an explanation on the origin of isolated O stars and massive O star clusters observed in the Milky Way.
\item We compare the distribution and mass function of dense cores in the two young O-star candidate clusters, RCW38 and W43, obtained with ALMA with the present results.
The two regions show signatures of cloud-cloud collisions as shown by \citet{2016ApJ...820...26F} and \citet{2020arXiv200110693K}.
The separation between O stars formed is typically 0.03 -- 0.1\,pc as determined by the dense core distribution in the collision-compressed layer.
They also show top-heavy core mass functions, which are consistent wth the present results, while more samples are desirable for better confrontations with observations.
\end{enumerate}

\begin{ack}
This work was supported by JSPS KAKENHI Grant Number 15H05694. 
We thank the anonymous referee for their careful reading of our manuscript and their many constructive comments.
Numerical computations were carried out on XT4 and XC30 system at the Center for Computational Astrophysics (CfCA) of National Astronomical Observatory of Japan.
\end{ack}

\bibliography{reference}

\newpage

\clearpage

\begin{table}
\tbl{Model parameters }{%
\begin{tabular}{lr@{\,}l}  
\hline\noalign{\vskip3pt} 
\multicolumn{1}{c}{Parameter} & \multicolumn{2}{c}{Value} \\  [2pt] 
\hline\noalign{\vskip3pt} 
$\langle n\rangle_{0}$ & 300 & cm$^{-3}$ \\
$\Delta n/\langle n\rangle_{0}$ & 0.33 & \\
$B_{0}$ & 20 & $\mu$G \\
$V_\mathrm{coll}$ & 10 & km\,s$^{-1}$ \\
Resolution & (8.0/512) & pc \\
\hline\noalign{\vskip3pt} 
\end{tabular}}\label{tab:modelparameters}
\end{table}

\begin{table}
\tbl{Mass of dense gas}{%
\begin{tabular}{rrrr}  
\hline\noalign{\vskip3pt} 
\multicolumn{1}{c}{$t$ (Myr)} & \multicolumn{3}{c}{Mass ($M_{\solar}$)} \\  [2pt] 
\cline{2-4}
 & $n>10^{4}$\,cm$^{-3}$ & $n>10^{5}$\,cm$^{-3}$ & $n>10^{6}$\,cm$^{-3}$ \\
\hline\noalign{\vskip3pt} 
0.2 & 2800 & \multicolumn{1}{c}{---} & \multicolumn{1}{c}{---} \\
0.3 &  4100 &   31 & \multicolumn{1}{c}{---} \\
0.4 &  5700 &  310 & 5 \\
0.5 &  9400 &  970 & 3 \\
0.6 & 13000 & 1800 &  38 \\
0.7 & 18000 & 2900 & 140 \\
\hline\noalign{\vskip3pt} 
\end{tabular}}\label{tab:mass}
\end{table}


\end{document}